\input harvmac

\skip0=\baselineskip
\divide\skip0 by 2
\def\tmpsp{\the\skip0}

\def\skipthis#1{{}}

\def\IR{\relax{\rm I\kern-.18em R}}
\def\IZ{\relax\ifmmode\hbox{Z\kern-.4em Z}\else{Z\kern-.4em Z}\fi}
\def\IQ{\relax{\rm I\kern-.40em Q}}
\def\IS{\relax{\rm I\kern-.18em S}}

\Title{\vbox{\baselineskip12pt\hbox{hep-th/0008083}
\hbox{HUTP-00/A032}}}
{\vbox{\centerline{Casimir Force in Compact Noncommutative}
        \vskip2pt\centerline{Extra Dimensions and Radius Stabilization}}}
\centerline{\bf Soonkeon Nam\foot{Permanent Address :
Dept. of Physics, Kyung Hee University; Seoul, 130-701, Korea,
{\tt nam@khu.ac.kr}}
} 
\bigskip\centerline{Department of Physics}
\centerline{Harvard University}
\centerline{Cambridge, MA 02138}
\centerline{\tt nam@pauli.harvard.edu}

\vskip .3in \centerline{\bf Abstract} 
We compute the one loop Casimir energy of an interacting 
scalar field in a compact noncommutative space
of $R^{1,d}\times T^2_\theta$, where we have ordinary flat
$1+d$ dimensional Minkowski space and two dimensional noncommuative torus.
We find that next order correction due to the noncommutativity still
contributes an attractive force and thus will have a quantum instability.
However, the case of vector field in a periodic boundary condition gives 
repulsive force for $d>5$ and we expect a stabilized radius.
This suggests a stabilization mechanism for a senario in Kaluza-Klein 
theory, where some of the extra dimensions are noncommutative. 
\smallskip

\Date{08/00}
\lref\kaluza{Th. Kaluza, Situngsber. d. K. Preuss. Akad. d. Wissen.
z. Berlin, Phys.-Math. Klasse (1921) 966.} 
\lref\klein{O. Klein, Z. F. Physik 37 (1926) 895.}
\lref\wolfram{J. Ambjorn and S. Wolfram, Ann. Phys. {\bf 147} (1983) 1.}
\lref\casimir{H.B.G. Casimir,
{\it ``On the Attraction Between Two Perfectly Conducting Plates,''}
{Proc. K. Ned. Akad. Wet.} {\bf 51} (1948) 793.}
\lref\noncomkk{J. Gomis, T. Mehen and M.B. Wise, { hep-th/0006160}.}
\lref\mrs{S. Minwalla, M.V. Raamsdonk and N. Seiberg, {\it``Noncommutative
Perturbative Dynamics''}, { hep-th/9912072}.}%
\lref\mvrs{M.V. Raamsdonk and N. Seiberg, {\it `` Comments on Noncommutative
Perturbative Dynamics''}, hep-th/0002186.}%
\lref\haya{M. Hayakawa,{\it ``Perturbative analysis on infrared
and ultraviolet aspects of
noncommutative QED on $R^4$,''} hep-th/9912167.}%
\lref\texa{W. Fischler, E. Gorbatov, A. Kashani-Poor, S. Paban,
P. Pouliot and J. Gomis, ``Evidence for winding states in
noncommutative quantum field theory,'' hep-th/0002067.}%

\lref\sunew{A. Matusis, L. Susskind and N. Toumbas,
``The IR/UV connection in the non-commutative gauge theories,''
hep-th/0002075.}%

An old idea, viewed in a new light, can have a new meaning. 
One of the oldest ideas of unifying gravity with other interactions is that
of Kaluza\kaluza\ and Klein\klein, i.e. that of  extra dimensions. 
Its modern rebirth came about with the advent of 
supergravity theories, and it is an
essential element in developments of string/M-theory.
\lref\largeextradim{
L. Randall and R. Sundrum, Phys. Rev. Lett. {\bf 83} (1999) 3370
[hep-ph/9905221];
Phys. Rev. Lett. {\bf 83} (1999) 4690 
[hep-th/9906064];
N. Arkani-Hamed, S. Dimopoulos, G. Dvali, and N. Kaloper, 
Phys. Rev. Lett. {\bf 84} (2000) 586 [hep-th/9907209].
}
\lref\expkk{
See for example T.G. Rizzo, {\it ``Indirect Collider Tests for Large
Extra Dimensions'',} hep-ph/9910255, and references therein.
}
\lref\namnet{Soonkeon Nam, JHEP {\bf 03} (2000) 005 [hep-th/9911104].}
\lref\namkk{Soonkeon Nam, JHEP {\bf 04} (2000) 002 [hep-th/9911237].}
\lref\kaloperkk{N. Kaloper, Phys. Lett. {\bf B474} (2000) 269 [hep-th/9912125].}

More recently, new possibilities of large extra dimension\largeextradim\
make the Kaluza-Klein modes something to be sought after in experiments
\expkk.
For example, in some models\namnet\ with more than two extra dimensions, there
is a distinct mass gap in Kaluza-Klein spectrum\namkk\kaloperkk.
So this eighty-years-old idea is still very much alive and is 
more so these days.

One question which is essential in Kaluza-Klein theories is how we can have
small (or unobserved) extra dimensions.
\lref\appelchodos{T. Appelquist and A. Chodos, Phys. Rev. Lett.
{\bf 50} (1983) 141; Phys. Rev. {\bf D28} (1983) 772.} 
To explain smallness of extra dimensions, Appelquist and 
Chodos\appelchodos\ suggested
that the vacuum fluctuations of the higher dimensional gravitational field
might provide a physical mechanism.
They considered linearized quantum gravity in $D$-dimensions and computed
the effective potential to one loop.
For the effective potential,
they obtained an infinite 
constant part\foot{
This infinity can of course be removed by proper dimensional
regularization.},
which is an induced cosmological
constant and an attractive part. They computed the one loop vacuum energy
in the compact extra dimensions, i.e. the gravitational 
Casimir energy\casimir.
Since the attractiveness of the Casimir energy pushed the size of the extra
dimension down to the Planck scale, the natural cutoff scale of the
linearized gravity, the {\it hope} was that presumably the dynamics of
Planck scale, where the nonperturbative quantum gravity sets in, will
stabilize the size of the extra dimensions.
\lref\rubinroth{M.A. Rubin and B.D. Roth, Nucl. Phys. 
{\bf B226} (1983) 444.}
Thermal effects could not introduce any stabilty: either the size is pushed
down to zero or to the infinity\rubinroth.
\lref\candelasweinberg{P. Candelas and S. Weinberg, 
Nucl. Phys. {\bf B237} (1984) 397;
A. Chodos and E. Meyers, Ann. Phys.  {\bf 156} (1984) 412.}
A very large number of light matter fields (around $10^{4-5}$)
can be introduced to stabilize the radius\candelasweinberg, since the
gravitational contribution per degree of freedom to the Casimir energy is
much larger than the matter contribution.

It is very natural to expect that there will be stabilization of the
size of the extra dimensions if there is some intrinsic {\it minimum}
length scale in the theory. One candidate certainly is the Planck 
(or string) scale as mentioned above. 
In this paper we will explore another possibility, when there is
noncommutativity of space in the extra dimensions.
When we have spacetime noncommutativity,
Lorentz invariance is broken. 
However, having extra dimensions with broken 
(or deformed) Lorentz symmetry is not incompatible with observations so far.

\lref\twoloopscalar{I. Ya Arefeva, D.M. Belov, and A.S. Koshelev,
Phys. Lett. {\bf B476} (2000) 431 [hep-th/9912075]; 
{\it ``A Note on UV/IR for Noncommutative Complex Scalar Field,''} 
hep-th/0001215; 
{\it ``UV/IR Mixing for Noncommutative Complex Scalar Field Theory, II (Interaction with Gauge Fields)''}
hep-th/0003176.}
\lref\ncqed{M. Hayakawa,
{\it ``Perturbative analysis on infrared and ultraviolet aspects of
noncommutative QED on $R^4$,} hep-th/9912167;
A. Matusis, L. Susskind, and N. Toumbas, 
{\it ``The IR/UV connection in the noncommutative gauge theories,''}
hep-th/0002075}
\lref\ncother{C.P. Martin, D. Sanchez-Ruiz, Phys. Rev. Lett. {\bf  83} 
(1999) 476 [hep-th/9903077];
M. Sheikh-Jabbari, JHEP {\bf 06} (1999) 015 [hep-th/9903107].
}
\lref\cds{A. Connes, M.R. Douglas, and A. Schwarz, JHEP {\bf 9802} (1998) 003
[hep-th/9711162].}

There has been a lot of attention recently on quantum field theories on 
noncommutative spaces\cds. 
Interacting scalar field theories\mrs\twoloopscalar, 
QED\ncqed\ and other theories were\ncother\ considered. 
\lref\ncsw{N. Seiberg and E. Witten,
JHEP {\bf 9909} (1999) 032 [hep-th/9908142].}
This class of field theories is very interesting, because it arises
naturally in the context of string theory\ncsw, 
and is a consistent theory by itself.
Here  we have the following commutation relations among
space-time coordinates $x^\mu$ 
\eqn\coomm{[x^\mu , x^\nu] = i\theta^{\mu\nu},}
and $\theta^{\mu\nu}$ introduces 
{\it minimum area} in the $\mu, \nu$ plane, just
as there is a minimum volume in phase space due to 
$[x,p]=i \hbar$,
due to a space-time uncertainty relations
\eqn\uncertainty
{\Delta x^\mu \Delta x^\nu \geq {1\over 2} |\theta^{\mu\nu}|.}
So in a noncommutative space there will be a length scale associated with
$\sqrt{|\theta^{\mu\nu}|}$.
Another consequence of this relation is the UV/IR mixing, due to the
absence of decoupling of scales. Short distance scales in one direction 
is related to long distance scales in another direction which is related to 
the previous one by the parameters $\theta^{\mu\nu}$.
\lref\greene{B.R. Greene, K. Schalm, and G. Shiu
{\it ``Warped Compactifications in M and F Theory''},
hep-th/0004103 .}

Difference with the Planck scale is that this scale is something which
should be determined by underlying dynamics. In this sense we will not be
able to solve the problem of radius stabilization in Kaluza-Klein
theories completely. 
However, in most compactification senarios in string theory, 
$B_{\mu\nu}$ has expectation value\greene, and 
we will be able to relate the
expectation value of $B_{\mu\nu}$ with the radius of the Kaluza Klein radius.

This is interesting because,
despite all the theoretical interests, the relevance 
of quantum field theories in  noncommuatative space to measurable 
effects in particle physics has not been discussed very much.
One of the main reason is that the presence of the external magnetic field
which induces the noncommutativity breaks Lorentz invariance of the
spacetime and thus a strong noncommutativity might not be a desirable thing 
to have.
However, having a noncommutative extra dimension can be interesting, without
destroying the desirable four dimensional Lorentz invariance.
\lref\nckk{J. Gomis, T. Mehen, and M.B. Wise, 
``Quantum Field Theories with Compact Noncommutative Extra Dimensions'',
hep-th/0006160.}
\lref\winding{W. Fischler, E. Gorbatov, A. Kashani-Poor, S. Paban,
P. Pouliot, and J. Gomis,
{\it ``Evidence for winding states in
noncommutative quantum field theory,''} hep-th/0002067.}
Very recently, there has been a work by Gomis et al\nckk\ in this direction 
where the Kaluza-Klein spectrum due to noncommutative compact extra
dimension has been considered. They obtain corrections to the Kaluza-Klein
spectrum which resembles the contributions of winding states in string
theory. 
\lref\stringncft{Y. Kiem and Sangmin Lee, 
{\it ``UV/IR Mixing in Noncommutative Field Theory via Open String Loops'',}
hep-th/0003145;
A. Bilal, Chong-Sun Chu, and R. Russo,
{\it ``String Theory and Noncommutative Field Theories at One Loop}'',
 hep-th/0003180.}
This is consistent with the 
close relations found between
string theory calculations and field theory calculations
in noncommutative space \stringncft\winding.

\lref\chodos{See for example Chap. 7 of T. Appelquist, A. Chodos, and 
P.G.O. Freund,``{\it Modern Kaluza-Klein Theories}'', Addison-Wesley, 
Menlo Park,  1987.}

\lref\bookcasimir{For the general introduction to Casimir force, 
see for example,
G. Pluien, B. M\"uller, and W. Greiner, Phys. Rep. {\bf 134} (1986)
87;
V.M. Mostepanenko and N.N. Trunov, 
{\it ``The Casimir Effect and its Applications,''}
Oxford Univ. Press, 1997.}
\lref\milton{
K.A. Milton, {\it ``The Casimir Effect: Physical Manifestation of the 
Zero-Point Energy,''} hep-th/9901011.}

In this paper, we explore  possible effect that the noncommutativity
in the extra dimensions might have on the Casimir force\bookcasimir\milton.
What we expect is that noncommutativity naturally introduces
a minimum volume, that of the Moyal cell,  which is proportional
to the noncommutativity parameter and will eventually stabilize the radius.
Therefore it is quite natural to expect that it will compete
with the attractive Casimir force.
In this paper, we find an interesting result that actually depending on the
type of the field, scalar or vector, the next order correction due to the
noncommutativity will be either attractive or repulsive.  The theory with
scalar field with $phi^3$ interaction 
has quantum instability whereas a theory with vector field stabilizes.

To be more quantitative, 
let us begin by a simple review of the 
derivation of the Casimir effect, when a massless
scalar field is confined between two parallel plates separated by a
distance $a$\casimir\wolfram.
(Similar analysis can certainly be done for massive fields, with similar
but more complicated results\wolfram.)
\lref\robin{A. Romeo and A.A. Saharian,
{\it ``Casimir effect for scalar fields under Robin boundary conditions on
plates,''} hep-th/0007242.}
Although Dirichlet,  Neumann,
or in general  Robin (i.e. mixed)\robin boundary condition
can be dealt with, we consider the Dirichlet case for simplicity.
\eqn\Dboundary{\phi(0) = \phi(a)=0.}
The Casimir force between the plates is obtained by summing up the
zero-point energy per unit area,
\eqn\Casienergy{
E/A = {1\over 2} \sum_{\alpha}  \omega_{\alpha} 
= {1\over 2} \sum_{n=1}^{\infty}\int {d^2 k \over (2\pi)^2} \sqrt
{k^2 + {n^2\pi^2\over a^2}},}
where for convenience we have set $\hbar = c =1$.
The integers $n=1,2, \cdots $ label the normal modes between the plates
and $k$ is the transverse momenta along the plates.
The sum as it stands is formally divergent. 

To extract a finite value from this divergent sum, 
we  invoke dimensional regularization.
Using the definition of the Gamma function,
\eqn\gammafunction{\Gamma (z) = p^z \int^{\infty}_{0} e^{-pt}t^{z-1} dt,}
we can put the energy per unit volume $A$ (at the boundary) 
$E/A$ as follows:
\eqn\energydensity{
E/A={1\over 2}\mu^{3-d}
 \sum_{n=1}^{\infty}\int {d^d k\over (2\pi)^d}
\int^{\infty}_{0} {dt\over t} t^{-1/2} e^{-t(k^2+n^2\pi^2/a^2)}{1\over 
\Gamma(-{1\over 2})}.}
In the above we have introduced an arbitrary mass scale $\mu$ to keep above
expression a four dimensional energy density.
From now on we will suppress the dependences on $\mu$.
Now we interchange the integrations over $t$ and $k$ and 
integrate over the transverse momenta, which is the Gaussian integration.
The final result is 
\eqn\finalresult{
E/A = -{1\over a^{d+1}}
\Gamma \left({d+2\over 2}\right)(4\pi )^{-(d+2)/2}\zeta(d+2).}
In obtaining this equation we have used the following reflection formula of 
Riemann zeta function to avoid infinities from $\Gamma(z)$ when
$z$ is a negative even integer:
\eqn\reflection
{\Gamma\left({z\over 2}\right)
\pi^{-z/2} \zeta(z) =
\Gamma\left({1-z\over 2}\right)
\pi^{(z-1)/2} \zeta(1-z).}
Since the energy is always is negative and falls off as $a$ decreases,
we always have attractive force between the plates due to the massless scalar
field.

Let us now generalize this to the noncommutative
$\phi^3$ field theory on $R^{1,d}\times T^2_{\theta}$.
By this we mean that
\eqn\nctorus
{[x^{d+1} , x^{d+2}] = i \theta,}
and other commutation relations among space time coordinates vanish. 
Also $0 \leq x^{d+1}, x^{d+2} \leq 2\pi R$.
We must have
an interacting theory to see the effects of the noncommutativity.
In a perturbative quantum field theory in noncommutative space, 
all the information about the noncommutativity can be put in the 
interaction vertices of Feynman diagrams.
Moreover, the effect of noncommutativity appears through nonplanar diagrams.

The theory we are considering is defined by the following action\mrs:
\eqn\action{S = \int d^{d+3}x
\left({1\over 2} (\partial \phi)^2 -
{1\over 2} m^2 \phi^2 - {\lambda\over 3!} \phi\star \phi\star\phi
                          \right).
}
The $\star$-product is defined by
\eqn\starproduct{
(f\star g)(x) = 
e^{{i\over 2} \theta^{\mu\nu}
{\partial \over \partial \alpha^\mu  }
{\partial \over \partial \beta^\mu  }}
f(x+\alpha)g(x+\beta)|_{\alpha =\beta=0},}
and introduces an infinite number of higher derivative interactions.
The kinetic part is not modified with the noncommutativity.

One can calculate the one loop contributions to the two point functions and
obtain the one loop corrections to the dispersion relation for nonzero
modes in the Kaluza-Klein spectrum\nckk,
when we have periodic boundary condition.
The Kaluza-Klein spectrum at one loop is as follows:
\eqn\kkmass{m^2_{\vec{n}}=
m^2+{\vec{n}^2\over R^2}-
{\lambda^2\over (4\pi)^3} \left(
{R^2\over {\vec{n}}^2\theta^2}   
+{5 \over 24} m^2 \ln \left( {m^2 \theta^2 {\vec n^2} \over  R^2 } 
\right)
+\cdots\right).  }
In the above, $\vec{n} = (n_{d+1}, n_{d+2})$ are integers
which give the quantized momenta 
$\vec{p} = \vec{n}/R$,
along the compact directions.
From now on we will simply put $\vec{n} = (n_1, n_2)$.
This mass formula resembles that of winding states in string theory.
The mass correction is negative, and for small values of $\theta$ 
the mass spectrum becomes tachyonic.
This certainly is not a healthy sign. Either there is an intrinsic 
instability in the theory or the perturbative analysis is not adequate
for small $\theta$'s.
In this paper, we will stay in the region where the perturbative analysis
is valid. We will only consider the case of massless scalar field, for the
sake of simplicity.

Casimir energy is again obtained by summing up all the modes,
to obtain the energy density in $d+1$ dimensions as follows:
\eqn\Casienergync{
u = E/A = {1\over 2} \sum_{\alpha} \omega_\alpha 
= {1\over 2} \sum_{n_1,n_2=1}^{\infty}\int {d^d \vec{k} \over (2\pi)^3} \sqrt
{k^2 + {\vec{n}^2\over R^2} - 
{\lambda^2 R^2\over (4\pi)^3\theta^2 \vec{n}^2 }}.
}
For the square root we can introduce the Schwinger's proper time
representation:
\eqn\nced{ u = {1\over 2} \sum_{n_1,n_2} 
\int {d^d k\over (2\pi)^d }
\int^\infty_0 {dt\over t} t^{-1/2}
e^{-t \left(\vec{k}^2 +\vec{n}^2/R^2 - 
\lambda^2 R^2/(4\pi)^3\theta^2\vec{n}^2\right)}.}
Now we integrate the transverse momentum $k$, by
doing a Gaussian integral, 
\eqn\ncedgi{ u = {-1\over 4 \sqrt{\pi}}
{1\over (4\pi   )^{d/2}} 
\sum_{n_1,n_2} 
\int^\infty_0 {dt\over t} t^{-(d+1)/2}
e^{-t \left(\vec{n}^2/R^2 - \lambda^2 R^2/(4\pi)^3\theta^2 \vec{n}^2\right)}.}
Again we  perform the $t$ integration using the integral representation of
Gamma function  and obtain
\eqn\ncedgii{ u = {-1\over 4 \sqrt{\pi}}
{1\over (4\pi   )^{d/2}} 
\Gamma\left(- {d+1\over 2 }\right)
\sum_{n_1,n_2} 
\left({{\vec{n}}^2\over R^2}-{\lambda^2 R^2\over
 (4\pi)^3\theta^2 {\vec{n}}^2} 
\right)^{(d+1)/2}.}
The infinite sum as it stands does not admit a closed form.
However, if we consider only the perturbative regime, then
we can expand in terms of 
$\lambda^2 a^2 /\theta^2  \ll 1,$ and have the following:
\eqn\ncedgiii
{\eqalign{u &= {-1\over 4 \sqrt{\pi}}
{1\over (4\pi   )^{d/2}} 
\Gamma\left(- {d+1\over 2 }\right)
\sum_{n_1,n_2} 
\left({{\vec{n}}^2\over R^2}\right)^{(d+1)/2}
\left[1
 - {d+1\over 2} {\lambda^2 R^4 \over (4\pi)^3\theta^2 
(\vec{n}^2)^2 }+\cdots \right],\cr
&=
{-1\over 2R^{d+1}}{1\over (4\pi)^{(d+1)/2}} 
\Gamma\left(- {d+1\over 2 }\right)
\left[
v_2\left(-d-1\over 2\right)- {d+1\over 2} 
{\lambda^2 R^4 \over (4\pi)^3\theta^2}
v_2\left(-d+3\over 2\right) 
+\cdots \right].\cr
}}
\lref\hardy{G.H. Hardy, Mess. Math. {\bf 49} (1919) 85.}
\lref\glasser{M.L. Glasser, J. Math. Phys. {\bf 14} (1973) 409.}
In the above we have used the following notation
and a relation found by Hardy\hardy:
\eqn\vacumf{v_2(s)\equiv \sum_{m,n=1}^\infty (m^2 + n^2 )^{-s}= 
\zeta(s) \beta(s) -\zeta(2s),}
where  $\beta(z)$ is defined as follows\glasser:
\eqn\betafunction{
\beta(z)  = \sum_{n=0}^\infty (-1)^n (2n+1)^{-z}.}
The value of this function is between $1/2$ and 1 for real positive $z$'s
and oscillates with unbound amplitude for negative real values of $z$.
Note that the Gamma function in the above function is divergent for odd
$d$'s.
\lref\zucker{I.J. Zucker, J. Phys. A. Math. Gen.
{\bf 9} (1974) 499.}
We can  use the `reflection formula'\zucker:
\eqn\reflect
{\Gamma(z)\zeta(z)\beta(z)\pi^{-z} = \Gamma(1-z)\zeta(1-z)\beta(1-z)\pi^{z-1}.}


So the structure of the energy per unit area is
\eqn\ff{ u = -\alpha/R^{d+1} - \gamma/R^{d-3}.}
where both $\alpha$ and $\gamma$ are both positive constants.
$\alpha$ is the coefficients one gets in the commutative case, and 
the ratio $\gamma/\alpha$ is 
\eqn\ratio{{\gamma\over \alpha}=
{-1 \over d-1} {\pi\lambda^2\over 8 \theta^2} 
{\Gamma\left({d-1\over2}\right)\zeta\left({d-1\over2}\right)
\beta\left({d-1\over2}\right) - 
\Gamma\left({d-2\over2}\right)\zeta(d-2)\sqrt{\pi}
\over 
\Gamma\left({d+3\over2}\right)\zeta\left({d+3\over2}\right)
\beta\left({d+3\over2}\right) - 
\Gamma\left({d+1\over2}\right)\zeta(d+1)\sqrt{\pi}}.}

We see that the contribution from the noncommutative part will never become 
repulsive and will not stabilize  the size of radis $R$ even for $d>3$.
The case of $d=3$ the contribution from the correction to the
Casimir energy constant, so there is no repulsive force.
So in the case of $d=3$, we see that up to the order of perturbation we
have used, there is no stabilization. So we 
might say that we have to consider the next order in correction to the
Casimir energy, i.e. higher order terms in Eq. 18.
Of course, in order to be really consistent  we first need the result for 
two loop self energy and it is beyond the scope of this paper.

In order to discuss the higher dimensional tori, each direction 
having different radius,  we can perform a similar calculation.
Again the vacuum energy is
\eqn\energy{
{1\over 2} \sum^\infty_{n=1} \int {d^d k\over  (2\pi)^3}
\sqrt{\vec{k}^2 + 
\sum_i\left(
{{{n_i}}^2 \over R^2_i} - {\rho^2}
{R^2_i \over {n^2_i}\theta^2  }+\cdots\right)}.}
In the above we have indicated the next order correction 
on dimensional ground the next order correction.

\lref\epstein{
P. Epstein, Math. Ann. {\bf 56} (1903) 516; {\bf 63} (1907) 205.
}
For this we need Epstein Zeta function\epstein\
\eqn\epstein{Z_p(1/a_1,\cdots,1/a_p ;s) 
= \sum_{n_1 =-\infty}^\infty \cdots
\sum_{n_p=-\infty}^{\infty }{'}
\left[ \left( {n_1 \over  a_1 } \right)^2
+  \cdots + \left( {n_p  \over  a_p }  \right)^2
      \right]^{-s/2}.}
Here the prime denotes that the term for which all $n_i=0$ is to be
omitted.
The qualitative feature will be similar, in the sense that there will a
attractive contribution, in the noncommutative limit, and the
noncommutative part will have attractive (and sometimes marginal)
contribution. 
This can be seen in the limit where the effects of the nonplanar part 
is maximal.
Of course for a quantitative result we have to resort to numerial methods.
We have seen that the noncommutative extra dimensions
cannot be stabilized with scalar fields 
with the introduction of the noncommutativity.

First of all this result should be generalized for the case of vector
and linearized gravity.
Since in a noncommutative spacetime a pure $U(1)$ gauge theory is 
{\it interacting} unlike in ordinary space, we will be able to see the
effects of noncommutativity in this theory.
Consider the action
\eqn\uone{
S= -{1\over 4} \int d^{d+3}x F_{MN}\star F^{MN},}
where the field strength is
\eqn\fieldstr{
F_{MN} =\partial_M A_N -\partial_N A_M
-ig (A_M \star A_N -A_N\star A_M),}
Actually the Kaluza-Klein spectrum from a noncommutative $U(1)$
gauge theory in six dimensions is available\nckk, and 
is given by the following formula:
\eqn\emkk{
m^2_{\vec{n}} =  {{\vec{n}^2}\over R^2} -
{8g^2 R^4\over \pi^3\theta^4(\vec{n}^2)^2}+ \cdots .}
There is a similarity with the scalar field case, but also a
difference. The dependence on the radius of the extra dimension
$R$ is different and has a higher power, and this will affect
the vacuum energy.

Following a similar analysis we have 
\eqn\ncedgiiv
{\eqalign{u &= {-2\over 4 \sqrt{\pi}}
{1\over (4\pi   )^{d/2}} 
\Gamma\left(- {d+1\over 2 }\right)
\sum_{n_1,n_2} 
\left({{\vec{n}}^2\over R^2}\right)^{(d+1)/2}
\left[1
 - {d+1\over 2} {8 g^2 R^6 \over \pi^3\theta^4 
(\vec{n}^2)^3 }+\cdots \right],\cr
&=
{-1\over  R^{d+1}}{1\over (4\pi)^{(d+1)/2}} 
\Gamma\left(- {d+1\over 2 }\right)
\left[
v_2\left(-d-1\over 2\right)- {d+1\over 2} 
{8g^2 R^6 \over \pi^3\theta^4}
v_2\left(-d+5\over 2\right) 
+\cdots \right].\cr
}}
We have multiplied the polarization factor of 2.
A rigorous proof that the case of electromagnetic field fluctuations
give the same result as the scalar case with a factor of two can be 
given\milton.
Since the ratio ${ v_2((-d+5)/2)\over v_2((-d-1)/2)}$ stays positive for all 
values of $d>5$ we now expect that the Casimir 
force will stabilize for $d>5$ and not for $d\leq 4$.
The value of the ratio decreases with increasing $d$. 
The compactification radius will be at 
\eqn\radius{
R=R_0= \left( 
\left(2\over d-5\right)
\left(\pi^3\theta^4\over 8 g^2\right)
\left({ v_2\left(-d+5\over 2\right)\over v_2\left(-d-1\over2\right)}\right)      \right)^{1\over6},}when there is only
a $U(1)$ vector field present.

This is to be contrasted to what we had for the scalar field.
It is expected that similar Kaluza-Klein spectrum for gravity will lead to
similar results.
Since the Casimir force for the gravity is far more dominant than that of
scalar fields, the field which is responsible for the compactifiction will
be the gravity field and we expect to have a stabilization of the size of
the extra dimensions.
Of course, it would be interesting to investigate these cases in detail.

\lref\horavafabinger{
M. Fabinger and P. Horava, Nucl. Phys. {\bf B580} (2000) 243 [hep-th/0002073].}
Recently, Casimir force between branes in  a flat $S^1/Z_2$ orbifold
compactification of M-theory was computed\horavafabinger.
Here one finds similar dependence in the distance between the branes, as in
the case with circle compactification\chodos.
It is expected that for the cases with more extra dimensions, similar
analysis as performed here will give a `stabilization' in the distances
between the branes.
\lref\radion{W.D. Goldberger and M.B. Wise, Phys. Rev. Lett. {\bf 83} 
(1999) 4922 [hep-ph/9907447].}
\lref\goldberger{W.D. Goldberger and I.Z. Rothstein, 
{\it Quantum Stabilization of Compactified $AdS_5$'',} hep-th/0007065.}
Another interesting work is by Goldberger and Rothstein\goldberger\  
on the quantum stabilization of radion stabilization\radion.
The system is such that it consists of two branes bounding a region of 
anti de Sitter space. 
It turns out that the quantum fluctuation destabilizes the system, just as
in the case of flat space.
It would be interesting to consider the consequences of noncommutativity in
the context of radion stabilization and study the quantum stability, which
we do expect.

\lref\susy{G.W. Gibbons and H. Nicolai,
Phys. Lett. {\bf 143B} (1984) 108;
T. Inami and K. Yamagishi, Phys. Lett. {\bf 143B} (1984) 115.} 
The Casimir effect in supergravity theories in a supersymmetric
backgrounds, have cancellation of the contribution from bosonic part by the
fermionic part\susy.
However, a finite temperature breaks supersymmetry and there will be a
finite Casimir effect in a senario of early universe.  

\centerline{\bf Acknowledgements}

I would like to thank C. Vafa and A. Strominger for the hospitality
at the string theory group of Harvard.
This work is supported in part by Brain Korea 21 (BK21)  
program of Korea Research Fund (2000), and by the Research Fund of Kyung Hee
University (2000).
I would like to thank C. Nunez and K. Olsen for discussions. 

\listrefs
\bye